\documentclass[superscriptaddress,showpacs,aps,prd,notitlepage]{revtex4-1}
\usepackage{epsfig}
\usepackage{amsmath,amssymb}
\usepackage{color}

\begin{document}
\title{Conformastationary disk-haloes  in Einstein-Maxwell gravity  II.  The physical interpretation of  the  halo}
\author{Antonio C. Guti\'errez-Pi\~{n}eres}
\email[e-mail:]{acgutierrez@correo.nucleares.unam.mx}
\affiliation{Instituto de Ciencias Nucleares, Universidad Nacional Aut\'onoma de M\'exico,
 \\AP 70543,  M\'exico, DF 04510, M\'exico}
\affiliation{Facultad de Ciencias B\'asicas,\\
Universidad Tecnol\'ogica de Bol\'ivar, Cartagena 13001, Colombia}

\author{Abra\~{a}o Capistrano}
\email[e-mail:]{abraao.capistrano@unila.edu.br}
\affiliation{Federal University of  Latin-American Integration, \\ Technological Park of Itaipu,
PO box 2123, Foz do Igua\c{c}u-PR 85867-670, Brazil}

\begin{abstract}
The relativistic  treatment  of  galaxies modelled as a rotating disk surrounded by  a  magnetised material  halo is  considered. The  galactic  halo is modelled by a  magnetised 
mass-energy distribution described  by  the  energy-momentum tensor of  a general fluid in  canonical form. All the  dynamics quantities characterising the  physical of  the  halo 
are  expressed in  exact form in terms  of  an arbitrary solution  of  the  Laplace's  equation. By way of illustration,  a  ``generalisation'' of the Kuzmin solution of the 
Laplace's equation is  used. The motion of  a  charged  particle on  the  halo region is  described. All the relevant quantities  and  the  motion  of  the charged  particle show a 
reasonable physical behaviour.
\end{abstract}

\maketitle

\section{introduction}
In the observational context, many ambiguities still exist  about the main constituents,  geometry and  dynamics (thermodynamics) of the disk-halos. However,  there  are several  
different observations  which  probe the  galactic and surrounding galactic magnetic  field. A current  revision  of  the status of our knowledge about the magnetic fields in our 
Milky Way and in nearby star-forming galaxies is summarised in \cite{beck2015magnetic}. Additionally, a study of  the   disk and halo rotation are reported  in \cite{de2014planck}, 
 whereas that in \cite{mikhailov2014magnetic} the possibility that magnetic fields can be generated in the outskirts of disks is  studied. Solutions for the Einstein  and 
Einstein-Maxwell Field Equations  which are consistently applicable in the context of astrophysical remains a topical problem. Nevertheless, the effects of magnetic fields on the 
physical processes in  galaxies and their disk-halo interaction have been scarcely considered in the past.  The presence of  the  electric  field on the dark matter halo models has 
been considered  in \cite{Chakraborty:2014paa} whereas  the  presence of  electromagnetic  field in the halo-disk  system has  been  studied in \cite{PhysRevD.87.044010, 
gutierrez2013variational}. In   the last mentioned works the gravitational sources are statics.

In a  precedent  paper \cite{Gutierrez-Pineres:2014tda}   we   considered  the conventional treatment of galaxies modelled as a  stationary (conformastationary) thin disk and,  
correspondingly,  we  associate the galactic halo  with the region  surrounding the disk. We  found  the  physical  quantities  characterising  the  disk  and  we  concluded that  
the  disk  is  made of  a   well-behavioured  general  relativistic  magnetised source. In the present work  we  are continuing our  research of the thin disk-halo systems  by 
studying  the physical  content  of  the  energy-momentum tensor  of  the  halo  obtained in such paper. As we  used the inverse  method  (where a solution of the field equations 
is taken and then the energy-momentum  tensor is obtained),  we  do not impose restriction   on  the physical properties of  the material constituting  the  halo.  Moreover, as   
it is  well  known that,   all  static spacetime  can  be  obtained from one  stationary, here we  generalise  the results for conformastatic  disk-halos obtained in  
\cite{PhysRevD.87.044010} for the special case when the electric potential vanishes. 

Our results  are  compatibles with the presented in \cite{Chakraborty:2014paa} on possible features of galactic  halo. Moreover, the description of the  motion of charged  
particles on disk here deduced  is  in agreement with the  results of analysis of particles motion in the  magnetised disks discussed  in \cite{garcia2014exact}. As far as we 
know, this  is  the  first relativistic  model describing analytically the  halo of  a rotating source in presence of  a magnetic  field.   To  study the  dynamics quantities 
characterising the  physics of  the  halo, in  Section \ref{sec:generalisedKS},  we  express these  in terms of  an arbitrary solution  of  the  Laplace's  equation and  then  we 
calculate these  for  “generalisation” of the Kuzmin solution of the Laplace’s equation.  We  describe the motion of  a  charged  particle on  the  halo region  in Section 
\ref{sec: Motion}.  Finally we complete the paper with a discussion of the results in  Section \ref{sec:conclude}.

\section{The halo of the generalised Kuzmin-like disks}\label{sec:generalisedKS}
\begin{figure}
$$\begin{array}{cc}
\epsfig{width=3.5in, file=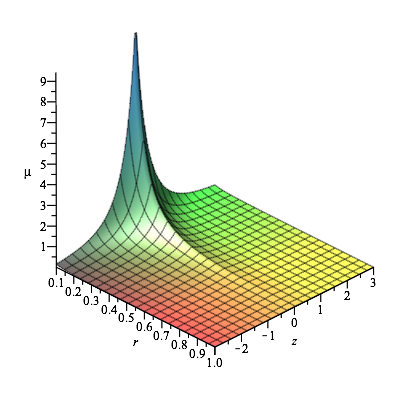} & 
\epsfig{width=3.5in, file=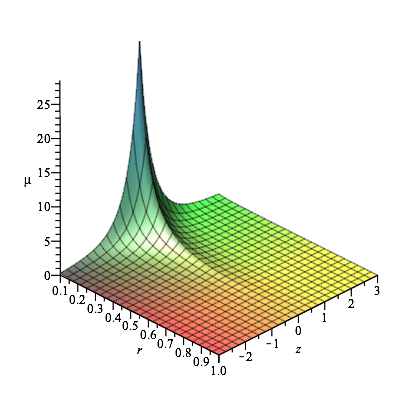} \\
 (a) & (b) \\
 \end{array}$$
\caption{\label{fig:figure1}Surface  plots    of  the  energy  density
 (a)  ${\mu}^{\pm}_{_0}$   and (b)  ${\mu}^{\pm}_{_1}$  on  the  exterior  halo   as a functions depending on  ${ r}$  and   ${z}$ with  parameters 
   $a={b}_0={b}_1= k=k_{\omega}=1$ and $\beta =0.75$}
\end{figure}

\begin{figure}
$$\begin{array}{cc}
\epsfig{width=3.5in, file=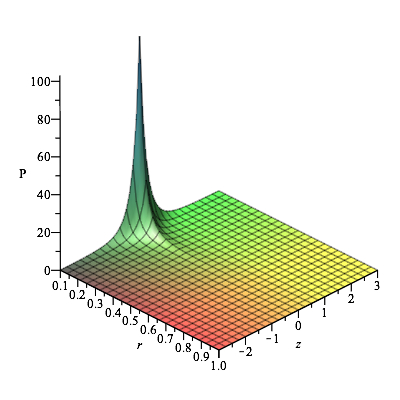} & 
\epsfig{width=3.5in, file=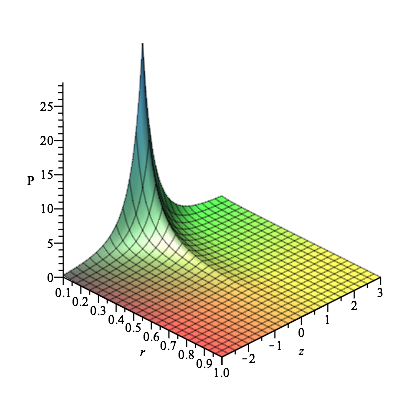} \\
 (a) & (b) \\
 \end{array}$$
\caption{\label{fig:figure2}Surface  plots    of  the  pressure
 (a)  ${P}^{\pm}_{_0}$   and (b)  ${P}^{\pm}_{_1}$  on  the  exterior  halo   as a functions depending on  ${ r}$  and   ${z}$ with  parameters 
   $a={b}_0={b}_1= k=k_{\omega}=1$ and $\beta =0.75$}
\end{figure}

In the  precedent  paper \cite{Gutierrez-Pineres:2014tda}  we   used  the  formalism presented in   \cite{PhysRevD.87.044010}  to obtain  an  exact relativistic  model   describing 
a  system composed  of  a  thin  disk surrounded by  a magnetised halo in a conformastationary space-time background.  However, although we   solved  the Einstein-Maxwell 
distributional field equations   to study the complete  system disk-halo,  we  only focused our  attention on the physical analysis of  the  disk.  In this  section, we will keep  
unchanged the the explicit  metric  and  magnetic  potentials in terms  of  the generalised  Kuzmin  solution of the Laplace's equation presented   in   
\cite{Gutierrez-Pineres:2014tda} and  we  calculate the principal physical  quantities  describing the  halo. To do so, we again  express  the 
energy-momentum tensor  of  the  halo   in the  canonical  form
        \begin{align}
          M_{\alpha\beta}^{\pm} = (\mu^{\pm} + P^{\pm})V_{\alpha}V_{\beta} + P^{\pm} g_{\alpha\beta}
                               + {\cal Q}_{\alpha}^{\pm}V_{\beta} + {\cal Q}_{\beta}^{\pm}V_{\alpha}
                               + \Pi_{\alpha\beta}^{\pm}\label{eq:EMTHcanonical}.
                 \end{align}
Consequently, we can  say that  the  halo is  constituted  by a some mass-energy distribution described by the last
energy-momentum  tensor   and $V^{\alpha}$  is the  four-velocity of certain observer. Accordingly $ \mu ^{\pm}$, $  P^{\pm}$, 
$ {\cal Q}_{\alpha}^{\pm}$ and $\Pi_{\alpha\beta}^{\pm}$ are then the  energy density, the isotropic  pressure,  the  heat flux and the
anisotropic  tensor on the  halo, respectively. Thus, it  is  immediate to see  that
                                   \begin{subequations}\begin{align}
                                      \mu ^{\pm}& = M_{\alpha\beta}^{\pm} V^{\alpha}V^{\beta},\label{eq:haloenergy}\\
                                             P^{\pm}& =\frac{1}{3}  {\cal H}^{\alpha\beta}M_{\alpha\beta}^{\pm},\label{eq:halopressure} \\
              {\cal Q}_{\alpha}^{\pm}& = -\mu ^{\pm} V_{\alpha} - M_{\alpha\beta}^{\pm} V^{\beta},\label{eq:haloheatflux}\\
             \Pi_{\alpha\beta}^{\pm}& ={\cal H}_{\alpha}^{\;\;\mu}{\cal H}_{\beta}^{\;\;\nu}  ( M_{\mu\nu}^{\pm}  - P^{\pm}{\cal H}_{\mu\nu}),\label{eq:haloanisotropict}
                                          \end{align}\label{eq:hobservables}\end{subequations}
with ${\cal H}_{\mu\nu}\equiv g_{\mu\nu} + V_{\mu}V_{\nu}$ and $\alpha=(t,r,\varphi)$. The  observer comoving with the fluid described by  the  energy-momentum  tensor  
(\ref{eq:EMTHcanonical}) will use  the  tetrad 
                          $\{ V^{\alpha},I^{\alpha},K^{\alpha},Y^{\alpha} \} \equiv     
                             \{h^{\;\;\; \alpha}_{(t)},h^{\;\;\; \alpha}_{(r)},h^{\;\;\; \alpha}_{(z)},h^{\;\;\; \alpha}_{(\varphi)}\}$, 
with  the  corresponding dual tetrad 
                        $\{V_{\alpha},I_{\alpha},K_{\alpha},Y_{\alpha} \} \equiv     
                           \{-h_{\;\;\; \alpha}^{(t)},h_{\;\;\; \alpha}^{(r)},h_{\;\;\; \alpha}^{(z)},h_{\;\;\; \alpha}^{(\varphi)}\}$,
where                                               
              \begin{subequations}\begin{eqnarray}
                {h}^{(t)}_{\quad \alpha}&=&F^{1/2}\{ 1,0,0, \omega \},\\
                {h}^{(r)}_{\quad\alpha}&=&F^{-\beta/2}\{ 0,1,0, 0 \},\\
                {h}^{(z)}_{\quad\alpha}&=&F^{-\beta/2}\{ 0,0,1,0 \},\\
                {h}^{(\varphi)}_{\quad\alpha} &=&F^{-\beta/2}\{ 0,0,0,r \}.
                         \end{eqnarray}\label{eq:localobservator}\end{subequations}             
The dual  vector     ${h}_{(\alpha)}^{\quad \beta}$    is  obtained by  the  condition $\eta_{(\alpha)(\beta)} = g_{\mu\nu} {h}_{(\alpha)}^{\quad \mu} {h}_{(\beta)}^{\quad \nu}$, 
where $F \equiv e^{2\phi}$  and $  e^{(1+\beta)\phi}= 1/(1-U)$, being $U$  a solution of  the Laplace's  equation. By using (\ref{eq:hobservables})  and (\ref{eq:EMTH})  we  obtain 
 for the  energy  density  and  the  pressure  of the  halo
                  \begin{align}
                        \mu ^{\pm} & = \frac{(U_{,r}^2+ U_{,z}^2)e^{2(1+2\beta)\phi}}{(1 + \beta)^2r^2}
                           \left\{ (2\beta+ \beta^2)r^2 - \frac{(1+ \beta)^2}{2k^2}r^2 e^{-2\phi} + \frac{3k_{\omega}^2(1+ \beta)^2}{4} \right\}
                            \end{align}
and
         \begin{align}
                      P^{\pm} & = \frac{(U_{,r}^2+ U_{,z}^2)e^{2(1+2\beta)\phi}}{3(1 + \beta)^2r^2}
                             \bigg\{ (4- 2\beta -  \beta^2)r^2  - \frac{(1+ \beta)^2}{2k^2}r^2 e^{-2\phi}
                              + \frac{k_{\omega}^2(1+ \beta)^2}{4} \left( 1 + 3k_{\omega}^2r^{-2}U^2 e^{2\beta\phi}(1 - e^{2\phi})  \right) \bigg\},
                         \end{align}
respectively. In  the same way,  by  inserting   (\ref{eq:EMTH}) into   (\ref{eq:hobservables})    we  obtain  for  the  heat flux of  the  halo
                \begin{align}
                  {\cal Q}_{\alpha}^{\pm}&= \frac{k_{\omega}e^{(1 + 2\beta)\phi}}{2(1 + \beta)r}
                                                                      \left\{2(1 + \beta)U_{,r} - (3 + \beta)r(U_{,r}^2  +  U_{,z}^2 )e^{(1+\beta)\phi} \right\}  
                                                                      \delta^{\varphi}_{\alpha},
                         \end{align}
moreover, it is easy  to  see  that   the  anisotropic tensor read
            \begin{align}
             \Pi_{\alpha\beta}^{\pm}  =     P_{r}^{\pm} I_{\alpha} I_{\beta} 
                                                                +  P_{z}^{\pm} K_{\alpha} K_{\beta}
                                                                +  P_{\varphi}^{\pm} Y_{\alpha} Y_{\beta} 
                                                                +  2P_{T}^{\pm}  I_{(\alpha} K_{\beta)}
                    \end{align}
where 
                \begin{align}
                       P_{r}^{\pm}&= e^{2\beta \phi} \Pi_{rr}^{\pm},\\
                       P_{z}^{\pm}&= e^{2\beta \phi} \Pi_{zz}^{\pm},\\
         P_{\varphi}^{\pm} &=  \frac{e^{2\beta \phi}}{r^2} \Pi_{\varphi\varphi}^{\pm},\\
                      P_{T}^{\pm}& = e^{2\beta \phi}    \Pi_{rz}^{\pm}.
                     \end{align}
and
        \begin{subequations}\begin{align}
         \Pi_{rr}^{\pm} &= \frac{e^{2(1+\beta)\phi}}{3(1 + \beta)^2r^2}
                                     \bigg \{
                                         \bigg(   \frac{k_{\omega}^2 (1 + \beta)^2}{2} + \frac{2(1+ \beta)^2 }{k^2}r^2e^{-2\phi} - 4 (1 + \beta - \beta^2)r^2
                                                -  \frac{3k_{\omega}^4 (1+ \beta)^2}{4}r^{-2}U^2 e^{2\beta\phi}(1 - e^{2\phi})     \bigg) U_r^2     \nonumber \\
                                    &  + \bigg(     -  k_{\omega}^2 (1 + \beta)^2          -  \frac{  (1+ \beta)^2 }{k^2}r^2e^{-2\phi}  +2 (1 + \beta - \beta^2)r^2
                                          -  \frac{3k_{\omega}^4 (1+ \beta)^2}{4}r^{-2}U^2 e^{2\beta\phi}(1 - e^{2\phi})  \bigg) U_z^2   \nonumber\\
                                    &   -   3(1 -\beta^2)r^2e^{-(1+ \beta)\phi}U_{,rr}
                                          \bigg \}, \\
        \Pi_{zz}^{\pm}&  =\frac{e^{2(1+\beta)\phi}}{3(1 + \beta)^2r^2}
                                      \bigg \{
                                           \bigg(          - {k_{\omega}^2 (1 + \beta)^2}     -   \frac{  (1+ \beta)^2 }{k^2}r^2e^{-2\phi} + 2  (1 + \beta - \beta^2)r^2
                                            -  \frac{3k_{\omega}^4 (1+ \beta)^2}{4}r^{-2}U^2 e^{2\beta\phi}(1 - e^{2\phi})  \bigg)U_r^2
                                                \nonumber\\
                                         &  + \bigg(      \frac{k_{\omega}^2 (1 + \beta)^2 }{2} + \frac{2(1+ \beta)^2 }{k^2}r^2e^{-2\phi}  - 4 (1 + \beta - \beta^2)r^2
                                          -  \frac{3k_{\omega}^4 (1+ \beta)^2}{4}r^{-2}U^2 e^{2\beta\phi}(1 - e^{2\phi}) \bigg)U_z^2
                                         \nonumber\\
                                       &  - 3(1 -\beta^2)r^2e^{-(1+ \beta)\phi}U_{,zz}
                                                 \bigg \}, \\
  \Pi_{\varphi\varphi}^{\pm}&= \frac{(U_{,r}^2+ U_{,z}^2)e^{2(1+\beta)\phi}}{3(1 + \beta)^2}
                                         \bigg\{
                                             2(1 +  \beta -  \beta^2)r^2  - \frac{(1+ \beta)^2}{k^2}r^2 e^{-2\phi}
                                            + \frac{k_{\omega}^2(1+ \beta)^2}{2} \left( 1 + 3k_{\omega}^2r^{-2}U^2 e^{2\beta\phi}(1 - e^{2\phi})  \right)
                                                \bigg\}
                                                \nonumber\\
                                      & - \frac{(1-\beta)}{1+ \beta}re^{(1+ \beta)\phi}U_{,r},\\
           \Pi_{rz}^{\pm} & =  \frac{e^{2(1 + \beta \phi)}}{(1 + \beta)^2}
                                \bigg\{
                          \bigg( -2(1 + \beta - \beta^2) + \frac{(1 + \beta)^2}{k^2}e^{-2\phi} +\frac{k_{\omega}^2(1 + \beta)^2}{2r^2}  \bigg) U_{,r}U_{,z}
                        - (1 -\beta^2) e^{-(1 + \beta)\phi} U_{,rz}               \bigg \}.
             \end{align}\label{eq:anisotropictensor}\end{subequations}
Notice that  ${\cal P}^{\pm} \equiv P_r^{\pm} + P_z^{\pm} + P_{\varphi}^{\pm}=0 $ and consequently the trace    $\Pi_{\quad\alpha}^{\pm \alpha} =0$. We have  obtained
 expressions for  the energy,  pressure and  the  another quantities characterising the dynamic  of the halo.   All  the dynamic quantities have been expressed  in 
terms of  an arbitrary $U(r,z)$  solutions of  the  Laplace's equation. Consequently, we have as many halo models as disk-like solutions of  the Laplace's equation. With  the  aim 
of  describe  the  halo surrounding the generalised Kuzmin-like disks presented  in  \cite{Gutierrez-Pineres:2014tda} we  
consider the  solution  of  the Laplace's  equation in the  form \cite{stephani2003exact}, \begin{eqnarray}
                    U= - \sum_{n=0}^{N}{\frac{b_n P_n(z/R)}{R^{n + 1}}} ,
                 \qquad P_n(z/R)= (-1)^n\frac{R^{n+1}}{n!}\frac{\partial^n}{\partial z^n}\left(\frac{1}{R}\right)
                    \label{eq:GeneralizedKuzmin},
                  \end{eqnarray}
 $P_n=P_n(z/R)$  being the  Legendre polynomials  in cylindrical  coordinates which  has been derived in the present form by a direct comparison of    the  Legendre polynomial 
expansion of   the  generating function with  a Taylor expansion of $1/r$, the radius $R$ denoted as $R^2\equiv r^2 + z^2$ and  $b_n$  arbitrary 
constant  coefficients. Thus, the  corresponding magnetic  potential   is
            \begin{eqnarray}
                  A_{\varphi} =- \frac{1}{k} \sum_{n=0}^{N} b_n \frac{(-1)^n}{n!}\label{eq:magpot}
                   \frac{\partial ^n}{\partial z^n}\left(\frac{z}{R}\right)
                  \end{eqnarray}
where  we  have imposed  $A_{\varphi}(0,z)=0$ in order  to  preserve the regularity  on the axis of  symmetry, and,  to  introduce  the corresponding  discontinuity  
in the  first-order derivatives of  the metric potential and  the magnetic  potential required  to  define the disk we  preform  the  transformation $z \rightarrow |z| +  a$. 
To  illustrate the results,  we  consider the two first members   $(N=0,1)$ of  the  family of the generalised Kuzmin-like disks (as shown in \ref{eq:GeneralizedKuzmin}). 
 In Fig. \ref{fig:figure1}, we show the behaviour of   energy  densities  ${{\mu}^{\pm}}$ on  the halo  as a function of $r$ and $z$.  In each case, we plot 
${\mu}^{\pm}_0(r,z)$  (Fig. \ref{fig:figure1}(a))  and ${\mu}^{\pm}_1(r,z)$  (Fig. \ref{fig:figure1}(b)) for the indicate values of the parameters. It can be seen that the 
surface energy density is everywhere positive and it vanishes sufficiently fast as $r$ increases.
In Fig. \ref{fig:figure2}, we show the behaviour of   pressure  ${P^{\pm}}$ on  the halo  as a function of $r$ and $z$.  In each case, we plot 
${P}^{\pm}_0(r,z)$ (Fig. \ref{fig:figure2}(a))  and ${P}^{\pm}_1(r,z)$  (Fig. \ref{fig:figure2}(b)) for the indicate values of the parameters.
We can see that pressure are always positive and behave as the energy density of the halo. We can see that the  behaviour of  these quantities described here are in agreement with 
the results published in \cite{Chakraborty:2014paa}. We also computed these functions for other values of the parameters within the
allowed range and in all cases we have found a similar behaviour.

\section{Motion of  a charged test particle in the halo}\label{sec: Motion}
\begin{figure}
$$\begin{array}{cc}
\epsfig{width=3.3in, file=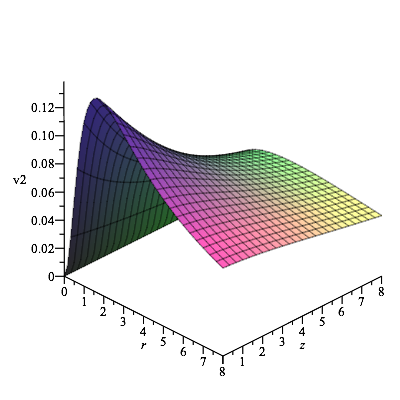} & 
\epsfig{width=2.8in, file=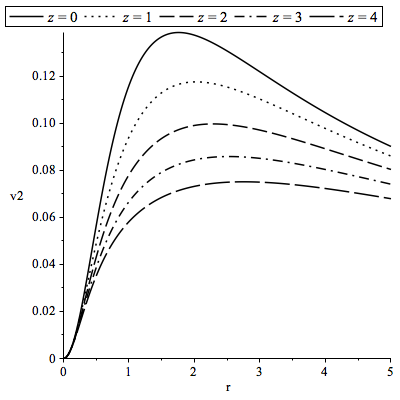} \\
 (a) & (b) \\
 \end{array}$$
\caption{\label{fig:figure3} Surface  plot    of  the  velocity (a)  ${v}^{ 2}_{_0}$   and $z$-slices   of the surface 
plot of the  velocity  (b)    on  the  exterior  halo   as a functions depending on  ${ r}$  and   ${z}$ 
with  parameters 
   $a={b}_0={b}_1= k=k_{\omega}=1$ and $\beta =0.75$}
\end{figure}

\begin{figure}
$$\begin{array}{cc}
\epsfig{width=3.3in, file=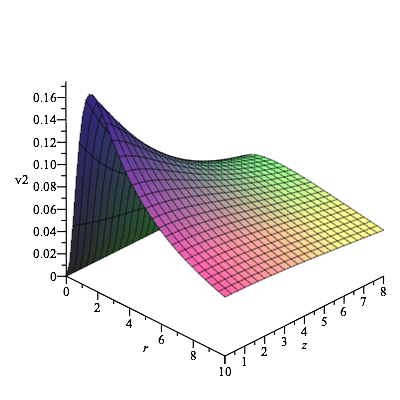} & 
\epsfig{width=2.8in, file=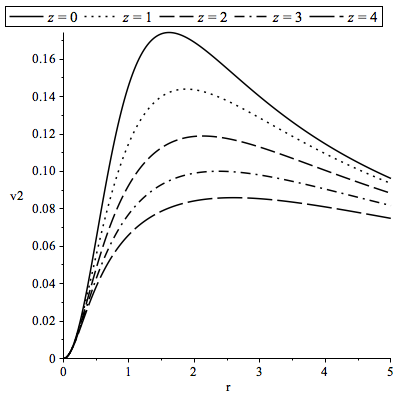} \\
 (a) & (b) \\
 \end{array}$$
\caption{\label{fig:figure4}Surface  plot    of  the  velocity
(a)  ${v}^{ 2}_{_1}$   and $z$-slices   of the surface plot of the  velocity (b) on the  exterior  halo   as a functions depending on  ${ r}$  and   ${z}$ with  parameters 
   $a={b}_0={b}_1= k=k_{\omega}=1$ and $\beta =0.75$}
\end{figure}

It  is interesting to  describe the  motion of  a  particle ``falling'' in the  halo, this kind of  motion is  called electrogeodesic.  Following 
\cite{landau1975classical}, the  equation of  motion  of  a  charged  particle  in a  gravitational and  electromagnetic fields (electrogeodesic equation)  is  obtained  by
                     \begin{align}
                         \frac{dv^{\alpha}}{ds} + \Gamma^{\alpha}_{\beta\gamma}v^{\beta}v^{\gamma}
                                                                        = \frac{e}{m}g^{\alpha\mu}F_{\mu\lambda}v^{\lambda},
                         \end{align}
where $e$ and $m$  are  the  charge  and  the mass of  the particle, respectively.  The   velocity of  the  particle as measured  by  the local observers 
 is  given  by   (see  Appendix B in \cite{Gutierrez-Pineres:2014tda} ) $v^{\alpha}=v^t(t^{\alpha }+ \Omega \varphi^{\alpha})$,  where
                         \begin{align}
                            v^t &=\frac{(1 -U)^{1/(1+ \beta)}}{(1 + k_{\omega}U\Omega)\sqrt{1 - v^2}}.
                         \end{align}
Here, the 3-velocity $v$ and  the  angular  velocity $\Omega$  of  the  particle as measured  by  the local observers  are given by
         \begin{align}
              v&=\frac{r \Omega (1 -U)}{1 + k_{\omega}U\Omega}
                      \end{align}
and              
             \begin{align}
              \Omega&=\frac{k_{\omega}(U_{,r}^2 + U_{,z}^2) \left((1 + \beta)   + \frac{2U}{1 -U}\right) \pm \sqrt{(U_{,r}^2 + U_{,z}^2) D}}
                                        {2(1 +  \beta)r (1 - U)^2U_{,r}  - 2 (U_{,r}^2 + U_{,z}^2)  A},\\
                       D&=  4(1 + \beta)r(1 - U)U_{,r} + (U_{,r}^2 + U_{,z}^2) \left( k_{\omega}^2(1 + \beta)^2   - 4\beta r^2     \right) ,\nonumber\\
                       A&= \beta r^2(1 - U) + k_{\omega}^2U \left( 1 + \beta + \frac{U}{1 -U}   \right),\nonumber
              \end{align}
respectively.   All the  quantities  depend on $r$ and $z$. In Fig. \ref{fig:figure3}(a) and Fig. \ref{fig:figure4}(a) we show the behaviour of  the velocity ${v^2}_0$  and  
${v^2}_1$ of  a charged  particle    following an electrogeodesic motion on  the halo  for  the  values of  indicated  parameters, respectively.  Additionally,  in  Fig. 
\ref{fig:figure3}(b) and Fig. \ref{fig:figure4}(b),   we  plot the $z$-slices   of the surface plot of the  velocity and ${v^2}_0$ and ${v^2}_1$  for the indicated values of the 
parameters, respectively. These  curves are obtained  via vertical slices  of the surface $v^2=v^2(r,z)$ (a vertical slice is a curve formed by the intersection of the surface 
$v^2=v^2(r,z)$ with the vertical planes).  For  each  curve,  we  can  see that the velocity is  always less than 1, its maximum  occurs around  $r=0$, and it vanishes sufficiently 
fast as $r$ increases. We also computed these functions for other values of the parameters within the allowed range and in all cases we found a similar behaviour. Naturally, the 
description of the  motion of charged  particles on disk  here deduced  is  in agreement with the  results of analysis of the electrogeodesic motion of the particle in the  
magnetised disks discussed  in \cite{garcia2014exact}.

\section{Concluding remarks} \label{sec:conclude}
In  the  present work we  continued our  research on  the  relativistic description of  the disk surrounded by  a halo   in presence of  an  electromagnetic field. As can be 
observed,  we   considered  the conventional treatment of galaxies modelled by  thin disk and,  correspondingly,  we  associate the galactic halo  with the region surrounding the 
disk. We   research started with  the  general formalism for a  conformastatic spacetime in  \cite{PhysRevD.87.044010} and then we generalized the formalism to the  
conformastationary  case in  \cite{Gutierrez-Pineres:2014tda}. Here, we  presented a physical description of  the  energy-momentum  tensor of  the halo.  We   concluded that  the  
disk  is  made of  a   well-behavioured  general  relativistic  source surrounded by a  well-behavioured  magnetized  halo ``material". 

As we  used the inverse  method,  no  restriction was  imposed  on  the physical properties of  the material constituting  the  halo.  In  fact, we expressed  the  energy-momentum 
tensor  of  the  halo in the canonical form.  The results obtained here  are all consistents  with the assumptions in the   precedent paper  \cite{Gutierrez-Pineres:2014tda} and  
generalise our  results presented  in  \cite{PhysRevD.87.044010}.   Accordingly,  when the  parameter $\beta$ in the metric  is equal to one the usual conformastationary line 
element is obtained and then the pressure and the anisotropic tensor on the  material constituting the  halo disappear. In a similar way, when  the parameter $k_{\omega}$ is equal 
to zero, the heat flux on the  halo vanishes, a feature of the static systems.  Furthermore, when we take simultaneously $k_{\omega}= 0$ and $\beta =1$, the results presented here 
describe the halo of  the  disks presented  in  \cite{PhysRevD.87.044010} for the special case when the electric potential vanishes. The  results presented here are compatibles 
wiht the  results in the  relativistic models of perfect fluid disks in a magnetic field presented  in \cite{garcia2014exact} and  with the  description of   galactic halo 
presented in \cite{Chakraborty:2014paa}.

To analyse the physical content of the energy-momentum tensor of  halo we expressed in the canonical form  and we projected it in a comoving frame defined trough of the local 
observers tetrad. Accordingly, we found the explicit expressions for the energy density, pressure,  heat flux, anisotropic tensor and electromagnetic potential of the halo  in 
terms a solution of the Laplace’s equation. Although  we used here  a “generalisation” of the Kuzmin solution of the Laplace’s equation, another models of disk-halos can be 
generated from  the solutions presented here through  suitable elections of  the  infinite family of solutions the Laplace’s equation.  Our  result presented here has  nothing to 
say  about dark matter  and exotic matter around of  the disk. We think that  its possible to use the  models presented here as  start point to generate  realistic models of 
relativistic  galaxies. 

\appendix
\section{The  energy-momentum tensor of  the  halo}\label{sec:EMTHexplicitly}
Following the  results presented in  \cite{Gutierrez-Pineres:2014tda},  f or  the metric
         \begin{eqnarray}
            ds^2= -e^{2\phi}(dt + \omega d\varphi)^2 + e^{-2\beta\phi} (dr^2 + dz^2 + r^2d\varphi^2),
            \quad e^{(1+\beta)\phi}=\frac{1}{1-U}, \quad\nabla^2 U=0,
             \label{eq:met1}
            \end{eqnarray}
the  non-zero components  of  the  energy-momentum  tensor of  the  halo  are
        \begin{subequations}\begin{align}
         M_{tt}^{\pm}&=\frac{U_{,r}^2 + U_{,z}^2}{(1 + \beta)^2}e^{4(1 + \beta)\phi}
                         \left\{ (\beta^2 + 2\beta) - \frac{1}{2k^2}(1 + \beta)^2 e^{-2\phi}
                            + \frac{3k_{\omega}^2}{4} (1+\beta)^2r^{-2} \right\}\\
   M_{t\varphi}^{\pm}&=\frac{k_{\omega}(U_{,r}^2 + U_{,z}^2)}{(1 + \beta)^2}e^{4(1 + \beta)\phi}
                         \left\{\frac{1}{2}(3 +\beta)(1 +\beta)e^{-(1 +\beta)\phi}
                          + U\left( \frac{3 k_{\omega}^2}{4}(1 + \beta)^2r^{-2} + \beta^2 + 2\beta\right)
                           \right\}\nonumber\\
                     &- k_{\omega}e^{2(1+\beta)\phi} \left( \frac{1}{2k^2}e^{2\beta\phi } U(U_{,r}^2 + U_{,z}^2) +
                        r^{-1}U_{,r},
                                  \right)\\
          M_{rr}^{\pm}&=- \frac{U_{,r}^2 - U_{,z}^2}{(1+\beta)^2}e^{2(1+\beta)\phi}
                          \left( (2 -\beta^2) - \frac{(1 + \beta)^2}{2k^2}e^{-2\phi} \right)
                      + \frac{(1-\beta)}{(1+ \beta)^2}e^{2(1+\beta)\phi}
                        \left( 2 U_{,r}^2 -(1 + \beta)e^{-(1 + \beta)\phi} U_{,rr} \right)\nonumber\\
                     &+ \frac{k_{\omega}^2}{4}r^{-2}e^{2(1+ \beta)\phi}(U_{,r}^2 - U_{,z}^2) \\
         M_{rz}^{\pm}& = -\frac{2U_{,r}U_{,z}}{(1+ \beta)^2}e^{2(1 + \beta)\phi}
                                            \left(   (2-\beta^2) - \frac{(1 + \beta)^2}{2k^2}e^{-2\phi}\right)
                                          + \frac{(1-\beta)}{(1+\beta)^2}e^{2(1+\beta)\phi}
                                          \left(2U_{,r}U_{,z} - (1 + \beta)e^{-(1+\beta)\phi }U_{,rz}
                                          \right)\nonumber\\
                                     & + \frac{1}{2}k_{\omega}^2r^{-2}e^{2(1+\beta)\phi} U_{,r}U_{,z},\\
       M_{zz}^{\pm}& = \frac{U_{,r}^2 - U_{,z}^2}{(1+\beta)^2}e^{2(1+\beta)\phi}
                                          \left(   (2-\beta^2) - \frac{(1 + \beta)^2}{2k^2}e^{-2\phi}\right)
                                      +  \frac{(1-\beta)}{(1+\beta)^2}e^{2(1+\beta)\phi}
                                         \left(  2 U_{,z}^2  - (1 + \beta)e^{-(1+ \beta)\phi} U_{,zz}   \right) \nonumber\\
                                 & - \frac{1}{4}k_{\omega}^2r^{-2}e^{2(1+\beta)\phi} (U_{,r}^2 - U_{,z}^2),\\
 M_{\varphi\varphi}^{\pm}& = \frac{r^2  (U_{,r}^2 + U_{,z}^2)}{(1+\beta)^2}
                                     e^{2(1+\beta)\phi} \left(   (2-\beta^2) - \frac{(1 + \beta)^2}{2k^2}e^{-2\phi}\right)
                                      - \frac{(1-\beta)r U_{,r}}{(1+\beta)} e^{(1+\beta)\phi} +k_{\omega}^2 e^{2(1+ \beta)\phi}{\cal K},\\
                     {\cal K} &=  \frac{ U_{,r}^2 + U_{,z}^2}{(1+\beta)^2}
                                      \left\{(\beta^2 + 2\beta)U^2 e^{2(1+\beta)\phi}+ (3+ \beta)(1+ \beta)Ue^{(1+\beta)\phi}
                                       + \frac{(1+ \beta)^2U^2 }{2}e^{2\beta\phi}\left(\frac{3}{2}k_{\omega}^2r^{-2} - \frac{1}{k^2}\right)
                                      + \frac{(1+ \beta)^2}{4} \right\}\nonumber\\
                                     &- 2r^{-1}UU_{,r},\nonumber
        \end{align}\label{eq:EMTH}\end{subequations}
  where all the  quantities depend on $r$ and $z$.


 %

       \end{document}